\documentstyle[12pt]{article}
\def\b{\rm b}
\def\g{\rm g}
\def\I{{\rm i}}
\def\inbar{\vrule height1.5ex width.4pt depth0pt}
\def\rlx{\relax\leavevmode}
\def\IC{\rlx\hbox{\,$\inbar\kern-.3em{\rm C}$}}
\def\non{ \nonumber }

\begin{document}

\newpage
\pagestyle{empty}
\begin{flushright}
DTP/94-53 \\
hep-th/9412192  \\
December 1994
\end{flushright}
\begin{center}
{\Large {\bf Boundary $K$-matrix for the quantum Mikhailov-Shabat model}}
\end{center}
\begin{center}
{\bf J. D. Kim\footnote{jideog.kim@durham.ac.uk} \footnote{On leave of absence
from Korea Advanced Institute of Science and Technology} }    \\
{\it Department of Mathematical Sciences, \\
University of Durham, Durham DH1 3LE, U.K. }  \\
\end{center}
\vspace{3cm}

\begin{center}
ABSTRACT
\end{center}

We present complete solutions of $K$-matrix
for the quantum Mikhailov-Shabat model.
It has been known that there are three diagonal solutions with no
free parameters, one being trivial identity solution, the others non-trivial.
The most general solutions which we found consist of three families
corresponding to each diagonal solutions.
One family of solutions depends on two arbitrary parameters.
If one of the parameters vanishes, the other must also vanish
so that the solutions reduces to trivial identity solution.
The other two families for each non-trivial diagonal solutions
have only one arbitrary parameter.

\newpage
\pagestyle{plain}

\section*{I. Introduction}
The Yang-Baxter equation whose solution is called $R$-matrix
appears frequently in physics, {\it eg.},
factorizable scattering theory on a line and solvable lattice models.
The actual adaptation of $R$-matrix to each of these theories requires
some more ingredients.

In the factorizable scattering theory, the solutions
of the Yang-Baxter equation can be used to compute the $S$-matrix, which
is given by $R$-matrix multiplied by overall factor chosen to satisfy crossing
relation and unitarity, {\it etc}. Here the spectral parameter is related
with coupling constant as well as the rapidity parameter of the particles
and the deformation parameter is related with coupling constant.
In the solvable lattice theory, the $L$-operator is given by $R$-matrix
and the bulk Hamiltonian can be computed from transfer matrix which is given
by the trace of products of $L$-operators.

Generalization of factorizable scattering theory to a half-line lead to
the discovery of the reflection equation
whose solution is called $K$-matrix\cite{Che}.
This equation describes interactions related with boundary.
By similar way as in the bulk theory, boundary $S$-matrix can be computed
from $K$-matrix\cite{GZ,Gho}.
This idea was also used to prove the quantum integrability of open spin chains
in the framework of quantum inverse scattering theory\cite{Skl}.
This gives us the Hamiltonian including boundary terms.

At the early stage of development, the solutions of the Yang-Baxter equation
were computed in a rather direct way, {\it ie.}, by solving the functional
equations component-wise. The substantial number of solutions were
compiled in Ref.\cite{KS} before the general algebraic solutions
were incidentally obtained\cite{Jim,Baz,KR}.

The most general non-diagonal solutions of reflection equation for
XXX, XXZ and XYZ type $R$-matrix which are $4\times4$ matrices were solved
in Ref.\cite{Che,Skl,DG,IKonno}.
The diagonal solutions for bigger size $R$-matrices were found
in Ref.\cite{MNR} for the Zamolodchikov-Fateev model\cite{ZF},
in Ref.\cite{MN} for the Mikhailov-Shabat model and
in Ref.\cite{DG} for $A_n^{(1)}$ model.
Spectral parameter independent solutions for various models are considered
in Ref.\cite{KSS}.

We consider the most general non-diagonal solutions of reflection equation
for the $9\times9$ $R$-matrix, especially corresponding to the quantum
Mikhailov-Shabat model\cite{IKorepin}.
This model is interesting because it is one of the relativistic quantum field
theory with single scalar field having non-trivial
families of boundary interaction\cite{Cor}.

This model was known to have two non-trivial diagonal solutions
of $K$-matrix as well as trivial identity solution\cite{MN}.
These solutions have no free parameters.
The most general solutions which we found consist of three families
corresponding to each diagonal solutions.
One family of solutions depends on two arbitrary parameters.
If one of the parameters vanishes, the other must also vanish
so that the solutions reduces to trivial identity solution.
The other two families for each non-trivial diagonal solutions
have only one arbitrary parameter and have upper-lower triangular structures.

The plan of this paper is as follows.
This introduction is the section I. In section II, we consider the whole
$9\times9$ equations resulting from matrix reflection equation and
describe their general features.
In section III, we solve the equations to obtain solutions. Finally,
we make some discussions in section IV.

\newpage
\section*{II. Reflection Equation}
The bulk $S$-matrix of integrable quantum field theories can be described
by the $R$-matrix which is the solution of the Yang-Baxter equation.
The $R$-matrix acts on $V \otimes V$, where $V$ is $\IC^3$
for the Mikhailov-Shabat model,
since this model has $3\times3$ Lax-pair representation
even though it involves only single scalar field.
\begin{equation}
R_{12}(u) R_{13}(u+v) R_{23}(v) = R_{23}(v) R_{13}(u+v) R_{12}(u),
\label{ybe}
\end{equation}
where $R_{12}(u)$, $R_{13}(u)$ and $R_{23}(u)$ act on $V \otimes
V \otimes V$, with $R_{12}(u)=R(u) \otimes 1$,
$R_{23}(u)=1\otimes R(u)$ {\it etc}.

The $R$-matrix for the Mikhailov-Shabat or
$A^{(2)}_2$  model was obtained in Ref.\cite{IKorepin} and
it was used to compute $S$-matrix in Ref.\cite{Smi}.
\begin{equation}
R(u)=\left(\begin{array}{ccc|ccc|ccc}
 c  &         &         &    &         &         &   &   & \\
    &    b    &         & e  &         &         &   &   & \\
    &         &    d    &    &    g    &         & f &   & \\ \hline
    & \bar{e} &         & b  &         &         &   &   & \\
    &         & \bar{g} &    &    a    &         & g &   & \\
    &         &         &    &         &   b     &   & e & \\ \hline
    &         & \bar{f} &    & \bar{g} &         & d &   & \\
    &         &         &    &         & \bar{e} &   & b & \\
    &         &         &    &         &         &   &   & c \end{array}
\right),
\end{equation}
where
\begin{eqnarray}
a(u) &=& -e^{u-3 \eta}+e^{-u+3 \eta}+e^{-\eta}-e^{\eta}+e^{-3 \eta}-e^{3
\eta}-e^{-5 \eta}+e^{ 5 \eta},  \non \\
b(u) &=& -e^{u - 3 \eta}+e^{-u + 3 \eta}+e^{-3 \eta}-e^{3 \eta},\non \\
c(u) &=& -e^{u - 5 \eta}+e^{-u + 5 \eta}+e^{-\eta}-e^{\eta},  \non \\
d(u) &=& -e^{u - \eta}+e^{-u + \eta}+e^{-\eta}-e^{\eta},  \\
e(u) &=&  e^{- \eta}-e^{-u + \eta}-e^{-5 \eta}+e^{-u +5 \eta}, \non \\
\bar{e}(u) & = & e^{u - \eta}-e^{\eta}-e^{u -5 \eta}+e^{5 \eta}, \non \\
f(u) & = & -e^{-u+\eta}+e^{-u-\eta}-e^{-u+3 \eta}+e^{3 \eta}+e^{-u+5 \eta}-e^{-
5 \eta},  \non \\
\bar{f}(u) & = & e^{u-\eta}-e^{u+\eta}+e^{u-3 \eta}-e^{-3 \eta}-e^{u-5
\eta}+e^{5 \eta}, \non \\
g(u) & = & -e^{-u } +1 +e^{-u + 4 \eta} -e^{4 \eta}, \non \\
\bar{g}(u) & = & -1 +e^{u} +e^{-4 \eta} -e^{u -4 \eta}.  \non
\end{eqnarray}
This $R$-matrix is a meromorphic function of $e^u$. It has only combined
$PT$ symmetry.
\begin{equation}
{\cal P}_{12} R_{12}(u) {\cal P}_{12} = R_{12}(u)^{t_1 t_2},
\end{equation}
where ${\cal P}_{ij}$ is the permutation operator on $V_i \otimes V_j$
defined by ${\cal P}(x \otimes y) = (y \otimes x)$
and $t_i$ denotes transposition in the $i$-th space.

The boundary versions of the Yang-Baxter equation so-called reflection
equations for $PT$ symmetric $R$-matrix are given by \cite{MN}
\begin{eqnarray}
\label{ref1}
\lefteqn{R_{12}(u-v) \stackrel{1}{K_-}(u) R_{12}^{t_1 t_2}(u+v)
  \stackrel{2}{K_-}(v) } \\
 & & = \stackrel{2}{K_-}(v) R_{12}(u+v) \stackrel{1}{K_-}(u) R_{12}^{t_1
t_2}(u-v), \non
\end{eqnarray}
\begin{eqnarray}
\label{ref2}
\lefteqn{R_{12}(-u+v) (\stackrel{1}{K_+})^{t_1}(u) \stackrel{1}{M^{-1}}
R_{12}^{t_1 t_2}(-u-v-2\rho) \stackrel{1}{M} (\stackrel{2}{K_+})^{t_2}(v) } \\
 & & = (\stackrel{2}{K_{+}})^{t_2}(v) \stackrel{1}{M} R_{12}(-u-v-2\rho)
\stackrel{1}{M^{-1}}  (\stackrel{1}{K_{+}})^{t_1}(u) R_{12}^{t_1 t_2}(-u+v),
\non
\end{eqnarray}
where $\stackrel{1}{K_-}(u)=K_{-}(u)\otimes 1$,
$\stackrel{2}{K_-}(u)=1\otimes K_{-}(u)$, { etc.} and
$M$ is given by $V$ matrix defined by the crossing unitarity relation.
\begin{eqnarray}
R_{12}(u) = \stackrel{1}{V} R_{12}^{t_2}(-u-\rho) \stackrel{1}{V^{-1}},
 & M=V^t V=M^t.
\end{eqnarray}
For $A_2^{(2)}$ model, $\rho=-6\eta-\I\pi$ and the matrix $V$ is given by
\begin{equation}
V=\left(\begin{array}{ccc}
             &         & -e^{-\eta}  \\
             &    1    &             \\
   -e^{\eta} &         &
\end{array}\right).
\end{equation}

In practice, if $K_{-}(u)$ is a solution of (\ref{ref1}) then
\begin{equation}
K_{+}(u)=K_{-}^t(-u-\rho) M
\label{auto}
\end{equation}
is a solution of (\ref{ref2}).

Now we write down the $9\times9$ component equations
in the following parametrization of $K$-matrix.
\begin{equation}
K(u,\eta)=\left(\begin{array}{ccc}
    \beta   &   x     & z \\
    \bar{x} & \alpha  & y \\
    \bar{z} & \bar{y} & \gamma \end{array}\right).
\label{Kmatrix}
\end{equation}
\begin{small}
\begin{displaymath}
\begin{array}{cl}
  2 & \rm{Unknowns}  \\
1133 :& z z c d = z z c d \\
1232 :& x x b \bar{g} + y x e d = y y b g + x y \bar{e} d  \\
3212 :& \bar{y} \bar{y} b g + \bar{x} \bar{y} \bar{e} d = \bar{x} \bar{x} b
\bar{g} + \bar{y} \bar{x} e d \\
3311 :& \bar{z} \bar{z} c d = \bar{z} \bar{z} c d  \\
  3 & \rm{Unknowns}   \\
1123 :& x z c b + z y c g = y z e g + z x e b + x z b d  \\
1132 :& z x c d = y z b g + z x b b + x z \bar{e} d  \\
1233 :& x z b \bar{g} + z y b b + y z e d = z y c d  \\
2133 :& x z \bar{e} \bar{g} + z y \bar{e} b + y z b d = y z c b + z x c \bar{g}
 \\
2311 :& \bar{x} \bar{z} c b + \bar{z} \bar{y} c g = \bar{y} \bar{z} e g +
\bar{z} \bar{x} e b + \bar{x} \bar{z} b d  \\
3211 :& \bar{z} \bar{x} c d = \bar{y} \bar{z} b g + \bar{z} \bar{x} b b +
\bar{x} \bar{z} \bar{e} d  \\
3312 :& \bar{x} \bar{z} b \bar{g} + \bar{z} \bar{y} b b + \bar{y} \bar{z} e d =
\bar{z} \bar{y} c d  \\
3321 :& \bar{y} \bar{z} c b + \bar{z} \bar{x} c \bar{g} = \bar{x} \bar{z}
\bar{e} \bar{g} + \bar{z} \bar{y} \bar{e} b + \bar{y} \bar{z} b d    \\
  4 & \rm{Unknowns}  \\
1111 :& x \bar{x} c e + z \bar{z} c f = \bar{x} x c e + \bar{z} z c f  \\
1212 :& x \bar{y} b g + \bar{x} x e c + y \bar{y} e f = \bar{x} y b g + x
\bar{x} e c + \bar{y} y e f  \\
2222 :& \bar{y} x g b + x \bar{y} \bar{g} b + \bar{x} x a \bar{e} + y \bar{y} a
e = y \bar{x} g b + \bar{x} y \bar{g} b + x \bar{x} a \bar{e} + \bar{y} y a e
\\
3232 :& \bar{y} x b \bar{g} + \bar{x} x \bar{e} \bar{f} + y \bar{y} \bar{e} c =
y \bar{x} b \bar{g} + x \bar{x} \bar{e} \bar{f} + \bar{y} y \bar{e} c  \\
3333 :& \bar{y} y c \bar{e} + \bar{z} z c \bar{f} = y \bar{y} c \bar{e} + z
\bar{z} c \bar{f}  \\
  5 & \rm{Unknowns}   \\
2123 :& x y \bar{e} a + \beta z \bar{e} \bar{e} + z \gamma \bar{e} e + y y b g
= y x e a + z \beta e \bar{e} + x x b \bar{g} + \gamma z e e  \\
2321 :& \bar{x} \bar{y} \bar{e} a + \beta \bar{z} \bar{e} \bar{e} + \bar{z}
\gamma \bar{e} e + \bar{y} \bar{y} b g = \bar{y} \bar{x} e a + \bar{z} \beta e
\bar{e} + \bar{x} \bar{x} b \bar{g} + \gamma \bar{z} e e  \\
  6 & \rm{Unknowns}   \\
1112 :& \bar{x} z b g+\alpha x e e +\beta x b b+x \beta e c+\bar{y} z e f =x
\alpha c e +\beta x c c +z \bar{y} c f  \\
1113 :& \alpha z g g+y x f e +x x g b+z \beta f c+\beta z d d +\gamma z f f=x y
c e+\beta z c c +z \gamma c f  \\
1121 :& \bar{x} z \bar{e} g+\alpha x b e +\beta x \bar{e} b+x \beta b c+\bar{y}
z b f =x \beta c b +z \bar{x} c g
\end{array}
\end{displaymath}
\begin{displaymath}
\begin{array}{cl}
1122 :& \alpha z a g+y x g e +x x a b+z \beta g c+\beta z \bar{g} d +\gamma z g
f=x x c b+z \alpha c g  \\
1131 :& \alpha z \bar{g} g+y x d e +x x \bar{g} b+z \beta d c+\beta z \bar{f} d
+\gamma z d f=z \beta c d  \\
1211 :& x \bar{z} b g+\alpha \bar{x} e e +\beta \bar{x} b b+\bar{x} \beta e c+y
\bar{z} e f =\bar{x} \alpha c e +\beta \bar{x} c c +\bar{z} y c f  \\
1223 :& x y b a+\alpha z e b +\beta z b \bar{e}+z \gamma b e=z \alpha e b +x y
b d  \\
1311 :& \alpha \bar{z} g g+\bar{y} \bar{x} f e +\bar{x} \bar{x} g b+\bar{z}
\beta f c+\beta \bar{z} d d +\gamma \bar{z} f f=\bar{x} \bar{y} c e+\beta
\bar{z} c c +\bar{z} \gamma c f  \\
1313 :& \bar{y} y f e+\bar{x} y g b +\bar{z} z f c =y \bar{y} f e+x \bar{y} g b
+z \bar{z} f c  \\
1333 :& \alpha z g \bar{g}+x y d \bar{e} +\beta z d \bar{f}+y y g b+\gamma z f
d +z \gamma d c=z \gamma c d  \\
2111 :& x \bar{z} \bar{e} g+\alpha \bar{x} b e +\beta \bar{x} \bar{e} b
+\bar{x} \beta b c +y \bar{z} b f=\bar{x} \beta c b +\bar{z} x c g  \\
2121 :& x \bar{x} \bar{e} a+z \bar{z} \bar{e} e +y \bar{x} b g=\bar{x} x
\bar{e} a+\bar{z} z \bar{e} e +\bar{y} x b g  \\
2132 :& y x b a+\alpha z \bar{e} b +z \beta b \bar{e}+\gamma z b e=z \alpha
\bar{e} b +y x b d  \\
2211 :& \alpha \bar{z} a g+\bar{y} \bar{x} g e +\bar{x} \bar{x} a b+\bar{z}
\beta g c+\beta \bar{z} \bar{g} d +\gamma \bar{z} g f=\bar{x} \bar{x} c
b+\bar{z} \alpha c g  \\
2233 :& \alpha z a \bar{g}+x y \bar{g} \bar{e} +y y a b+\beta z \bar{g}
\bar{f}+\gamma z g d +z \gamma \bar{g} c=y y c b+z \alpha c \bar{g}  \\
2312 :& \bar{x} \bar{y} b a+\alpha \bar{z} e b +\beta \bar{z} b \bar{e}+\bar{z}
\gamma b e=\bar{z} \alpha e b +\bar{x} \bar{y} b d  \\
2323 :& \bar{y} y e a+\bar{z} z e \bar{e} +\bar{x} y b \bar{g}=y \bar{y} e a+z
\bar{z} e \bar{e} +x \bar{y} b \bar{g}  \\
2333 :& \bar{y} z e \bar{g}+\alpha y b \bar{e} +\bar{x} z b \bar{f}+\gamma y e
b+y \gamma b c =y \gamma c b+z \bar{y} c \bar{g}  \\
3111 :& \alpha \bar{z} \bar{g} g+\bar{y} \bar{x} d e +\bar{x} \bar{x} \bar{g} b
+\bar{z} \beta d c+\beta \bar{z} \bar{f} d +\gamma \bar{z} d f=\bar{z} \beta c
d  \\
3131 :& x \bar{x} \bar{f} \bar{e}+y \bar{x} \bar{g} b +z \bar{z} \bar{f}
c=\bar{x} x \bar{f} \bar{e}+\bar{y} x \bar{g} b +\bar{z} z \bar{f} c  \\
3133 :& \alpha z \bar{g} \bar{g}+x y \bar{f} \bar{e} +\beta z \bar{f} \bar{f}+y
y \bar{g} b+\gamma z d d +z \gamma \bar{f} c=y x c \bar{e}+z \beta c \bar{f}
+\gamma z c c  \\
3221 :& \bar{y} \bar{x} b a+\alpha \bar{z} \bar{e} b +\bar{z} \beta b
\bar{e}+\gamma \bar{z} b e=\bar{z} \alpha \bar{e} b +\bar{y} \bar{x} b d  \\
3233 :& \bar{y} z b \bar{g}+\alpha y \bar{e} \bar{e} +\bar{x} z \bar{e}
\bar{f}+\gamma y b b+y \gamma \bar{e} c =y \alpha c \bar{e}+z \bar{x} c
\bar{f}+\gamma y c c  \\
3313 :& \alpha \bar{z} g \bar{g}+\bar{x} \bar{y} d \bar{e} +\bar{y} \bar{y} g
b+\beta \bar{z} d \bar{f}+\gamma \bar{z} f d +\bar{z} \gamma d c=\bar{z} \gamma
c d  \\
3322 :& \alpha \bar{z} a \bar{g}+\bar{x} \bar{y} \bar{g} \bar{e} +\bar{y}
\bar{y} a b+\beta \bar{z} \bar{g} \bar{f}+\gamma \bar{z} g d +\bar{z} \gamma
\bar{g} c=\bar{y} \bar{y} c b+\bar{z} \alpha c \bar{g}  \\
3323 :& y \bar{z} e \bar{g}+\alpha \bar{y} b \bar{e} +x \bar{z} b
\bar{f}+\gamma \bar{y} e b+\bar{y} \gamma b c =\bar{y} \gamma c b+\bar{z} y c
\bar{g}  \\
3331 :& \alpha \bar{z} \bar{g} \bar{g}+\bar{x} \bar{y} \bar{f} \bar{e} +\beta
\bar{z} \bar{f} \bar{f}+\bar{y} \bar{y} \bar{g} b +\gamma \bar{z} d d +\bar{z}
\gamma \bar{f} c =\bar{y} \bar{x} c \bar{e}+\bar{z} \beta c \bar{f}+\gamma
\bar{z} c c
\\
3332 :& y \bar{z} b \bar{g}+\alpha \bar{y} \bar{e} \bar{e} +x \bar{z} \bar{e}
\bar{f}+\gamma \bar{y} b b+\bar{y} \gamma \bar{e} c =\bar{y} \alpha c
\bar{e}+\bar{z} x c \bar{f}+\gamma \bar{y} c c  \\
  7 & \rm{Unknowns}   \\
1213 :& \alpha y g g+y \alpha f e +x \alpha g b+z \bar{x} f c+\beta y d d
+\gamma y f f \\
 & =x \gamma b g+\alpha y e e +\beta y b b+\bar{x} z e c+y \gamma e f  \\
1231 :& \alpha y \bar{g} g+y \alpha d e +x \alpha \bar{g} b+z \bar{x} d c+\beta
y \bar{f} d +\gamma y d f \\
 & =x \beta b \bar{g}+z \bar{x} b b +y \beta e d  \\
1312 :& \alpha \bar{y} g g+\bar{y} \alpha f e +\bar{x} \alpha g b+\bar{z} x f
c+\beta \bar{y} d d +\gamma \bar{y} f f \\
 & =\bar{x} \gamma b g+\alpha \bar{y} e e +\beta \bar{y} b b+x \bar{z} e
c+\bar{y} \gamma e f
\end{array}
\end{displaymath}
\begin{displaymath}
\begin{array}{cl}
1332 :& \alpha x g \bar{g}+x \alpha d \bar{e} +\beta x d \bar{f}+y \alpha g
b+\gamma x f d +z \bar{y} d c \\
 & =y \gamma b g+z \bar{y} b b +x \gamma \bar{e} d  \\
2212 :& \alpha \bar{y} a g +\bar{y} \alpha g e +\bar{x} \alpha a b+\bar{z} x g
c+\beta \bar{y} \bar{g} d +\gamma \bar{y} g f \\
 & =\bar{x} \alpha b a+\alpha \bar{x} e b +\beta \bar{x} b \bar{e}+\bar{z} y b
e+\bar{y} \alpha e g  \\
3112 :& \alpha \bar{y} \bar{g} g+\bar{y} \alpha d e +\bar{x} \alpha \bar{g}
b+\bar{z} x d c+\beta \bar{y} \bar{f} d +\gamma \bar{y} d f \\
 & =\bar{x} \beta b \bar{g}+\bar{z} x b b +\bar{y} \beta e d  \\
3132 :& \alpha x \bar{g} \bar{g}+x \alpha \bar{f} \bar{e} +\beta x \bar{f}
\bar{f}+y \alpha \bar{g} b+\gamma x d d +z \bar{y} \bar{f} c \\
 & =y \beta b \bar{g}+\alpha x \bar{e} \bar{e} +x \beta \bar{e} \bar{f}+\gamma
x b b+\bar{y} z \bar{e} c  \\
3213 :& \alpha \bar{x} g \bar{g}+\bar{x} \alpha d \bar{e} +\beta \bar{x} d
\bar{f}+\bar{y} \alpha g b+\gamma \bar{x} f d +\bar{z} y d c \\
 & =\bar{y} \gamma b g+\bar{z} y b b +\bar{x} \gamma \bar{e} d  \\
3231 :& \alpha \bar{x} \bar{g} \bar{g}+\bar{x} \alpha \bar{f} \bar{e} +\beta
\bar{x} \bar{f} \bar{f}+\bar{y} \alpha \bar{g} b+\gamma \bar{x} d d +\bar{z} y
\bar{f} c \\
 & =\bar{y} \beta b \bar{g}+\alpha \bar{x} \bar{e} \bar{e} +\bar{x} \beta
\bar{e} \bar{f}+\gamma \bar{x} b b+y \bar{z} \bar{e} c  \\
  8 & \rm{Unknowns}    \\
1221 :& \bar{x} y \bar{e} g+\alpha \alpha b e +\beta \alpha \bar{e} b+x \bar{x}
b c+\bar{y} y b f  \\
&=x \bar{x} b a+\alpha \beta e b+\beta \beta b \bar{e} +z \bar{z} b e+y \bar{x}
e g \\
1222 :& \alpha y a g+y \alpha g e +x \alpha a b+z \bar{x} g c+\beta y \bar{g} d
+\gamma y g f \\
 & =x \alpha b a+\alpha x e b +\beta x b \bar{e}+z \bar{y} b e+y \alpha e g  \\
1321 :& \alpha \bar{x} g a+\bar{y} \beta f b +x \bar{z} d b+\bar{x} \beta g
\bar{e}+\beta \bar{x} d \bar{g} +y \bar{z} g e+\gamma \bar{x} f g  \\
 & =\bar{x} \gamma \bar{e} g +\alpha \bar{y} b e+\beta \bar{y} \bar{e} b+x
\bar{z} b c +\bar{y} \gamma b f  \\
1323 :& \alpha y g a+\bar{y} z f b +x \gamma d b+\bar{x} z g \bar{e}+\beta y d
\bar{g} +y \gamma g e+\gamma y f g \\
 & =y \gamma e g +z \bar{y} e b+x \gamma b d  \\
2112 :& x \bar{y} \bar{e} g+\alpha \alpha b e +\beta \alpha \bar{e} b+\bar{x} x
b c+y \bar{y} b f  \\
 & =\bar{x} x b a+\alpha \beta e b+\beta \beta b \bar{e} +\bar{z} z b e+\bar{y}
x e g  \\
2113 :& \alpha x g a+y \beta f b +\bar{x} z d b+x \beta g \bar{e}+\beta x d
\bar{g} +\bar{y} z g e+\gamma x f g \\
 & =x \gamma \bar{e} g +\alpha y b e+\beta y \bar{e} b+\bar{x} z b c +y \gamma
b f  \\
2122 :& \alpha x a a+y \beta g b +\bar{x} z \bar{g} b+x \beta a \bar{e}+\beta x
\bar{g} \bar{g} +\bar{y} z a e+\gamma x g g \\
 & =x \alpha \bar{e} a +\alpha x b b+\beta x \bar{e} \bar{e}+z \bar{y} \bar{e}
e +y \alpha b g  \\
2131 :& \alpha x \bar{g} a+y \beta d b +\bar{x} z \bar{f} b+x \beta \bar{g}
\bar{e}+\beta x \bar{f} \bar{g} +\bar{y} z \bar{g} e+\gamma x d g \\
 & =x \beta \bar{e} \bar{g} +z \bar{x} \bar{e} b+y \beta b d  \\
2221 :& \alpha \bar{x} a a+\bar{y} \beta g b +x \bar{z} \bar{g} b+\bar{x} \beta
a \bar{e}+\beta \bar{x} \bar{g} \bar{g} +y \bar{z} a e+\gamma \bar{x} g g \\
 & =\bar{x} \alpha \bar{e} a +\alpha \bar{x} b b+\beta \bar{x} \bar{e}
\bar{e}+\bar{z} y \bar{e} e +\bar{y} \alpha b g  \\
2223 :& \alpha y a a+\bar{y} z g b +x \gamma \bar{g} b+\bar{x} z a
\bar{e}+\beta y \bar{g} \bar{g} +y \gamma a e+\gamma y g g \\
 & =y \alpha e a +\alpha y b b+z \bar{x} e \bar{e}+x \alpha b \bar{g} +\gamma y
e e
\end{array}
\end{displaymath}
\begin{displaymath}
\begin{array}{cl}
2231 :& \alpha \alpha \bar{g} a+y \bar{x} d b +\bar{x} y \bar{f} b+x \bar{x}
\bar{g} \bar{e}+\beta \alpha \bar{f} \bar{g}+\bar{y} y \bar{g} e +\gamma \alpha
d g \\
 & =\alpha \beta a \bar{g}+\beta \beta \bar{g} \bar{f} +y \bar{x} a b+\gamma
\beta g d+z \bar{z} \bar{g} c  \\
2232 :& \alpha x a \bar{g}+x \alpha \bar{g} \bar{e} +\beta x \bar{g} \bar{f}+y
\alpha a b+\gamma x g d +z \bar{y} \bar{g} c \\
 & =y \alpha b a+ \alpha y \bar{e} b +z \bar{x} b \bar{e}+x \alpha \bar{e}
\bar{g}+\gamma y b e  \\
2313 :& \alpha \bar{y} g a+y \bar{z} f b +\bar{x} \gamma d b+x \bar{z} g
\bar{e}+\beta \bar{y} d \bar{g} +\bar{y} \gamma g e+\gamma \bar{y} f g \\
 & =\bar{y} \gamma e g +\bar{z} y e b+\bar{x} \gamma b d  \\
2322 :& \alpha \bar{y} a a+y \bar{z} g b +\bar{x} \gamma \bar{g} b+x \bar{z} a
\bar{e}+\beta \bar{y} \bar{g} \bar{g} +\bar{y} \gamma a e+\gamma \bar{y} g g \\
 & =\bar{y} \alpha e a +\alpha \bar{y} b b+\bar{z} x e \bar{e}+\bar{x} \alpha b
\bar{g} +\gamma \bar{y} e e   \\
2331 :& \alpha \bar{y} \bar{g} a+y \bar{z} d b +\bar{x} \gamma \bar{f} b+x
\bar{z} \bar{g} \bar{e}+\beta \bar{y} \bar{f} \bar{g} +\bar{y} \gamma \bar{g}
e+\gamma \bar{y} d g \\
 & =\bar{y} \beta e \bar{g} +\alpha \bar{x} b \bar{e}+\bar{x} \beta b
\bar{f}+\gamma \bar{x} e b +y \bar{z} b c  \\
2332 :& \bar{y} x e \bar{g}+\alpha \alpha b \bar{e} +\bar{x} x b \bar{f}+\gamma
\alpha e b+y \bar{y} b c  \\
 & =y \bar{y} b a+\alpha \gamma \bar{e} b+z \bar{z} b \bar{e} +x \bar{y}
\bar{e} \bar{g}+\gamma \gamma b e  \\
3121 :& \alpha \bar{x} \bar{g} a+\bar{y} \beta d b +x \bar{z} \bar{f} b+\bar{x}
\beta \bar{g} \bar{e}+\beta \bar{x} \bar{f} \bar{g} +y \bar{z} \bar{g} e+\gamma
\bar{x} d g \\
 & =\bar{x} \beta \bar{e} \bar{g} +\bar{z} x \bar{e} b+\bar{y} \beta b d  \\
3122 :& \alpha \alpha \bar{g} a+\bar{y} x d b +x \bar{y} \bar{f} b+\beta \alpha
\bar{f} \bar{g}+y \bar{y} \bar{g} e +\gamma \alpha d g \\
 & =\alpha \beta a \bar{g}+\beta \beta \bar{g} \bar{f} +\bar{y} x a b+\gamma
\beta g d+\bar{z} z \bar{g} c  \\
3123 :& \alpha y \bar{g} a+\bar{y} z d b +x \gamma \bar{f} b+\bar{x} z \bar{g}
\bar{e}+\beta y \bar{f} \bar{g} +y \gamma \bar{g} e+\gamma y d g \\
 & =y \beta e \bar{g} +\alpha x b \bar{e}+x \beta b \bar{f}+\gamma x e b
+\bar{y} z b c
 \\
3222 :& \alpha \bar{x} a \bar{g}+\bar{x} \alpha \bar{g} \bar{e} +\beta \bar{x}
\bar{g} \bar{f}+\bar{y} \alpha a b+\gamma \bar{x} g d +\bar{z} y \bar{g} c \\
 & =\bar{y} \alpha b a+\alpha \bar{y} \bar{e} b +\bar{z} x b \bar{e}+\bar{x}
\alpha \bar{e} \bar{g}+\gamma \bar{y} b e  \\
3223 :& y \bar{x} e \bar{g}+\alpha \alpha b \bar{e} +x \bar{x} b \bar{f}+\gamma
\alpha e b+\bar{y} y b c  \\
 & =\bar{y} y b a+\alpha \gamma \bar{e} b+\bar{z} z b \bar{e} +\bar{x} y
\bar{e} \bar{g}+\gamma \gamma b e
\end{array}
\end{displaymath}
\begin{displaymath}
\begin{array}{cl}
  9 & \rm{Unknowns}   \\
1322 :& \alpha \alpha g a+\bar{y} x f b +x \bar{y} d b +\bar{x} x g
\bar{e}+\beta \alpha d \bar{g} +y \bar{y} g e+\gamma \alpha f g \\
 & =\alpha \gamma a g +y \bar{y} g e+x \bar{y} a b +z \bar{z} g c +\beta \gamma
\bar{g} d +\gamma \gamma g f  \\
1331 :& \alpha \beta g \bar{g}+x \bar{x} d \bar{e} +\beta \beta d \bar{f} +y
\bar{x} g b+\gamma \beta f d +z \bar{z} d c \\
 & =\alpha \gamma \bar{g} g+y \bar{y} d e +x \bar{y} \bar{g} b +z \bar{z} d c
+\beta \gamma \bar{f} d +\gamma \gamma d f  \\
2213 :& \alpha \alpha g a+y \bar{x} f b +\bar{x} y d b +x \bar{x} g
\bar{e}+\beta \alpha d \bar{g} +\bar{y} y g e+\gamma \alpha f g \\
 & =\alpha \gamma a g +\bar{y} y g e+\bar{x} y a b +\bar{z} z g c +\beta \gamma
\bar{g} d +\gamma \gamma g f  \\
3113 :& \alpha \beta g \bar{g}+\bar{x} x d \bar{e} +\beta \beta d \bar{f}
+\bar{y} x g b+\gamma \beta f d +\bar{z} z d c \\
 & =\alpha \gamma \bar{g} g+\bar{y} y d e +\bar{x} y \bar{g} b +\bar{z} z d c
+\beta \gamma \bar{f} d +\gamma \gamma d f
\end{array}
\end{displaymath}
\end{small}
It is understood that the first, second, third and fourth factors
in each terms have respectively the arguements $u,v,u-v$ and $u+v$.
Eqs.(1133, 3311) are automatically satisfied.
We numbered equations by $(ijkl)$, where $ij$ denotes
the rows and $kl$ represents the columns of the matrix reflection
equation in tensor notation.
Equations are sorted out according to the number of $\alpha, \beta, \gamma,
x, y, z, \bar{x}, \bar{y}$, and $\bar{z}$ variables.

Diagonal solutions for $K_{-}(u)$ have been obtained in \cite{MN}.
In this case, it is sufficient to solve Eqs.(1221,1331,1132) only.
It turns out that there are three solutions, being $K_{-}(u)=1$,
$K_{-}(u) = \Gamma^+(u)$ and $K_{-}(u) = \Gamma^-(u)$, with
\begin{equation}
\Gamma^{\pm}(u)=\left(\begin{array}{ccc}
  B_{\pm}(u)  &    &     \\
              & 1  &     \\
              &    & G_{\pm}(u)
\end{array} \right),
\label{diagonal}
\end{equation}
where
\begin{equation}
B_{\pm}= \frac{2 + \b_1^{\pm}(1-e^{-u})}{2 + b_1^{\pm}(1-e^{u})}, ~~~~
 \b_1^{\pm}=\frac{2~(-1\pm \I e^{-3 \eta})}{1+e^{-6\eta}},
\label{diagsol}
\end{equation}
\begin{displaymath}
G_{\pm}= \frac{2 - \g_1^{\pm}(1-e^{u})}{2 - g_1^{\pm}(1-e^{-u})}, ~~~~
 \g_1^{\pm}=\frac{2~(1\pm \I e^{3 \eta})}{1+e^{6\eta}}.
\end{displaymath}
These solutions have no free parameters. By the automorphism (\ref{auto}),
three solutions for $K_{+}(u)$ follow.

\newpage
\section*{III. Solutions}
Now we solve the $9\times9$ equations listed in the previous section.
The basic method of solving which we will use is the same as that
used when solving the Yang-Baxter equation in a direct way\cite{KS}.
The known solutions of $K$-matrix up to now are all obtained
by this method. The way is, first divide equations by $\alpha(u)$,
then take the derivative with respect to the variable $v$
and finally set $v$ to zero. This gives functional equations involving
the variable $u$ only among the elements of $K$-matrix, which we need to solve.

We suppose that the solutions to be proportional to identity
when the spectral parameter $u$ is zero.
\begin{equation}
X(0)=\bar{X}(0)=Y(0)=\bar{Y}(0)=Z(0)=\bar{Z}(0)=0,~~~~ B(0)=G(0)=1.
\end{equation}
We introduce the following functions normalized by $\alpha(u)$
\begin{eqnarray}
X(u) = \frac{x(u)} {\alpha(u)},~
Y(u) = \frac{y(u)} {\alpha(u)},~
Z(u) = \frac{ z(u)}{\alpha(u)},~
B(u) = \frac{ \beta(u)} {\alpha(u)},  \\
\bar{X}(u) = \frac{\bar{x}(u)} {\alpha(u)},~
\bar{Y}(u) = \frac{\bar{y}(u)} {\alpha(u)},~
\bar{Z}(u) = \frac{\bar{z}(u)} {\alpha(u)},~
G(u) = \frac{\gamma(u)}{\alpha(u)},  \non
\end{eqnarray}
and define their first derivatives at $u=0$ as follows.
\begin{eqnarray}
x_1 = X^{'}(u) |_{u=0}, &y_1 = Y^{'}(u) |_{u=0},& z_1 = Z^{'}(u) |_{u=0}, \\
\bar{x_1} = \bar{X}^{'}(u) |_{u=0}, & \bar{y_1} = \bar{Y}^{'}(u) |_{u=0}, &
\bar{z_1} = \bar{Z}^{'}(u) |_{u=0}, \non  \\
b_1= B^{'}(u) |_{u=0}, &  g_1=G^{'}(u) |_{u=0}. & \non
\end{eqnarray}

To begin with, let us suppose that all the elements of $K$-matrix
in Eq.(\ref{Kmatrix}) are non-zero and see what happens.
Following the procedure described above, we get relations between
$Y(u), X(u)$ from Eq.(1232) and $\bar{Y}(u), \bar{X}(u)$ from Eq.(3212)
for the 2 unknowns.
\begin{eqnarray}
Y(u) &=& \frac{-x_1 b \bar{g}+y_1 \bar{e} d}{x_1 e d-y_1 b g} X(u),  \\
 &=& \frac{e^{-\eta +u} (x_1 -e^u x_1 + e^{3 \eta} y_1 + e^{\eta + u} y_1) }
  {e^{2 \eta} x_1 + e^u x_1 - e^{3 \eta} y_1 + e^{3 \eta + u}  y_1 }
   X(u),  \non  \\
\bar{Y}(u) &=& \frac{-\bar{x_1} b \bar{g}+\bar{y_1} \bar{e} d}
                    {\bar{x_1} e d-\bar{y_1} b g} \bar{X}(u),  \\
 &=& \frac{e^{-\eta + u} (\bar{x_1} - e^u \bar{x_1} + e^{3 \eta} \bar{y_1}
      + e^{\eta + u} \bar{y_1} ) }
  {e^{2 \eta} \bar{x_1} + e^u \bar{x_1} - e^{3 \eta} \bar{y_1}
      + e^{3 \eta + u} \bar{y_1} } \bar{X}(u). \non
\end{eqnarray}
Next we go to the equations involving three unknowns. These equations
can be used to express $Z(u)$ and $\bar{Z}(u)$ in terms of $X(u)$
and $\bar{X}(u)$ respectively.
Eqs.(1123, 1132, 1233, 2133) give the same $Z(u)$ and
Eqs.(2311, 3211, 3312, 3321) give the same $\bar{Z}(u)$.
\begin{eqnarray}
Z(u) &=& {{\left( {e^{2 \eta}} + {e^{2 u}} \right)  { z_1} }
  \over {{e^{2 \eta}} { x_1} + {e^u} { x_1} - {e^{3 \eta}} { y_1} +
      {e^{3 \eta + u}} { y_1}}} X(u),  \\
\bar{Z}(u) &=&
  {{\left( {e^{2 \eta}} + {e^{2 u}} \right)  { \bar{z_1}} }
  \over {{e^{2 \eta}} { \bar{x_1}} + {e^u} { \bar{x_1}}
  - {e^{3 \eta}} { \bar{y_1}} + {e^{3 \eta + u}} { \bar{y_1}}}} \bar{X}(u).
\end{eqnarray}

Until now, the equations involved $X,Y,Z$ and their barred partners
seperately. However, the equations for 4 unknowns begin to connect the
unbarred and barred variables. We use these equations to determine
$\bar{X}(u)$ from $X(u)$. It turns out that five equations give the
same simple relations between $\bar{X}(u)$ and $X(u)$
\begin{equation}
\bar{X}(u)= \frac{\bar{x_1}}{x_1} X(u),
\end{equation}
if the following relations holds.
\begin{equation}
\bar{y_1} x_1=\bar{x_1} y_1.
\label{run4}
\end{equation}
Eq.(\ref{run4}) also give some simplifications to already obtained
relations.

Now we have relations among off-diagonal elements of $K$-matrix
which can be completely determined once we know $X(u)$. To determine
$X(u)$ we turn to equations involving 6 unknowns, skipping equations
involving 5 unknowns. There are sufficient number of equations for
6 unknowns to determine $\bar{X}(u)$. However, the answer is
rather unexpected in the sense that there is no solutions with
every elements of $K$-matrix non-vanishing. Let us see how this happens.

First we notice that from the paired equations like (1121, 2111)
we get the following relations.
\begin{equation}
\bar{z_1}=\frac{ \bar{x_1} \bar{x_1}} { x_1 x_1} z_1.
\end{equation}
Among 28 equations, we use Eqs.(1211, 2111)
to express $B(u)$ in terms of $X(u)$. Two equations give
two different expressions for the same $B(u)$ function.
\begin{equation}
B1(u)=\frac{1}{\bar{x_1} \bar{e} b} \left( -X \bar{z_1} \bar{e} g-\bar{x_1} b
e-Y \bar{z_1} b f+\bar{Z} x_1 c g+2 \bar{X} ((b,c)) \right),
\end{equation}
\begin{equation}
B2(u)=\frac{1}{\bar{x_1}(b b-c c)} \left( -X \bar{z_1} b g-\bar{x_1} e
e-\bar{X} b_1 e c-Y \bar{z_1} e f +\bar{Z} y_1 c f+2 \bar{X} ((e,c)) \right),
\end{equation}
where we introduced Wronskian of two functions as follows.
\begin{equation}
((f,g)) = f^{'} g-f g^{'}.
\end{equation}
Equating these two different expressions, we can get an expression for $X(u)$.
Similarly we use the equations (2333,3233) to get another expression
for $X(u)$ through $G(u)$ function.
\begin{equation}
G1(u)=\frac{1}{y_1 e b} \left( -\bar{Y} z_1 e \bar{g}-y_1 b \bar{e}-\bar{X} z_1
b \bar{f}+Z y_1 c \bar{g}+2 Y ((b,c)) \right),
\end{equation}
\begin{equation}
G2(u)=\frac{1}{y_1 (b b-c c)} \left( -\bar{Y} z_1 b \bar{g}-y_1 \bar{e}
\bar{e}-Y g_1 \bar{e} c-\bar{X} z_1 \bar{e} \bar{f} +Z \bar{x_1} c \bar{f}+2 Y
((\bar{e},c)) \right).
\end{equation}
So we have two different expressions for the same $X(u)$
function. Requiring these to be identical, we get
important information about the solutions. There are two possibilities.
\begin{equation}
\bar{x_1} ~z_1 =0, ~~y_1=e^{-\eta} x_1, ~~ b_1~ x_1=0, ~~ g_1 ~x_1=0.
\label{fullcond1}
\end{equation}
\begin{equation}
\bar{x_1} ~z_1 =0, ~~y_1=-e^{-\eta} x_1, ~~ b_1 =-\frac{4}{e^{2 \eta}+1},
{}~~ g_1 =\frac{4 e^{2 \eta} }{e^{2 \eta}+1}.
\label{fullcond2}
\end{equation}
It should be noted that the first derivatives of each elements of $K$-matrix
were assumed as non-zeros. So the conclusion is, no solution of $K$-matrix
with every components non-vanishing.

Now we consider the cases when some of the elements of $K$-matrix vanish.
When $\bar{x_1}$ is non-zero, above two conditions are still valid so
$z_1$ must be zero.
When $z_1=0$, the four equations(1131, 1223, 1333, 2132) imply $X(u)=Y(u)=0$.
So in this case, only $\bar{X}(u),\bar{Y}(u)$ and $\bar{Z}(u)$ can be non-zero.
The remaining possibility is only the case when $\bar{x_1}$ itself vanishes.
Eq.(3212) for 2 unknowns means $\bar{Y}$ must also vanish.
Moreover, we can easily see that Eqs.(2212,1312,3213,3231,3112) for 7 unknowns
imply either $\bar{Z}=0$ or that $X(u)=Y(x)=0$.
So in this case, there are two possibilities, non-zero $X(u),Y(u),Z(u)$
or non-zero $\bar{Z}, Z(u)$.

In summary, there are three possibilities. Case (I): non-zero
$Z(u),\bar{Z}(u)$,
Case (II): non-zero $X(u), Y(u), Z(u)$,
Case (III): non-zero $\bar{X}(u), \bar{Y}(u), \bar{Z}(u)$.
For any cases, $B(u),G(u)$ are non-zero. Let me first consider the case (I).

Case(I): It is easy to see that $\bar{Z}(u)$ should be proportional
to $Z(u)$ {\it eg.}, from Eq.(1111).
\begin{equation}
\bar{Z}(u)=\frac{ \bar{z_1} }{z_1} Z(u).
\end{equation}
It is useful to start from the Eqs.(2112,2332)
to express the $B(u)$ and $G(u)$ in terms of $Z(u)$.
They give
\begin{eqnarray}
\label{BI}
B(u) &=& \frac{2 ((e,b))-\bar{z_1} Z(u) b e-b_1 e b}
              {2 ((\bar{e},b)) + b_1 b \bar{e} }   \\
    &=& { {2 - \left( b_1+\bar{z}_1 Z(u) \right) \left( -1 + e^{-u} \right) }
 \over {2 +  b_1 \left( 1 - e^u \right) }} ,  \non
\end{eqnarray}
\begin{eqnarray}
\label{GI}
G(u) &=& \frac{2 ((\bar{e},b))-\bar{z_1} Z(u) b \bar{e}-g_1 \bar{e} b}
              {2 ((e,b))+g_1 b e }  \\
 &=& { {2 - \left( g_1+\bar{z}_1 Z(u) \right) \left( 1 - e^{u} \right) }
 \over {2 +  g_1 \left( e^{-u} -1 \right) }}. \non
\end{eqnarray}
 From Eq.(2123) for 5 unknowns, we get the following relation.
\begin{equation}
B(u) z_1 \bar{e} \bar{e}-G(u) z_1 e e=Z(u) (b_1-g_1) e \bar{e} +2 Z(u)
((\bar{e},e)).
\label{BGI}
\end{equation}
Inserting the expressions for $B(u)$ and $G(u)$ in Eqs.(\ref{BI},\ref{GI}) into
Eq.(\ref{BGI}), we can get an expression for $Z(u)$.
\begin{equation}
 Z(u) = {{2 \left( -1 + {e^u} \right)  \left( 1 + {e^u} \right)  { z_1}}
 \over{ZF}},
\end{equation}
where
\begin{small}
\begin{equation}
ZF =  4 {e^u} + 2 { b_1} ({e^u} - 2 {e^{2 u}}) + 2 { g_1}(1-{e^u})
    + { b_1} { g_1}(1  - 2 {e^u} + {e^{2 u}})
    + { \bar{z_1}} { z_1}( -1 + 2 {e^u} - {e^{2 u}}).
\end{equation}
\end{small}
Inserting back this into Eqs.(\ref{BI},\ref{GI}), we obtain $B(u)$ and $G(u)$
free of $Z(u)$. We determined the essential form of the solutions.

Now we have to see what constraints must be made on the parameters.
Only one equation is sufficient to determine it.
Let us use Eq.(2211) for 6 unknowns.
\begin{equation}
\bar{z}_1 a g+\bar{Z} b_1 g c- 2 \bar{Z}((g,c))
+B \bar{z}_1 \bar{g} d+G \bar{z}_1 g f=0.
\end{equation}
 From above equation, we can get
\begin{equation}
g_1=e^{2 \eta} b_1, ~~~ \bar{z}_1=e^{2 \eta} \frac{ b_1^2}{z_1}.
\end{equation}
In all, the solutions for the case (I) is the following.
\begin{eqnarray}
Z(u) &=&  {{ \left( -1 + {e^{2 u}} \right)  { z_1}}\over
    { 2 {e^u} + b_1 ( e^{2 \eta} + e^u - e^{2 u} -e^{2 \eta +u} ) }},  \\
\bar{Z}(u) &=& {{ e^{2 \eta} ( -1 + e^{2 u} ) {b_1}^2 } \over
 { (2 {e^u} +b_1 (e^{2 \eta} +e^u -e^{2 u} -e^{2 \eta +u} )) z_1 }}, \non \\
B(u) &=&  {{ 2 {e^u} +b_1 ( -1 + e^{2 \eta} +  e^u -e^{2 \eta + u} ) }\over
 { 2 {e^u} + b_1 ( e^{2 \eta} + e^u - e^{2 u} -e^{2 \eta +u} ) }}, \non  \\
G(u) &=& {{ e^u ( 2 + b_1 (1 - e^{2 \eta} - e^u + e^{2 \eta + u} )) }\over
 { 2 {e^u} + b_1 ( e^{2 \eta} + e^u - e^{2 u} -e^{2 \eta +u} ) }}. \non
\end{eqnarray}
$b_1$ and $z_1$ are the two free parameters.
If $b_1$ is zero, $g_1,~z_1,~\bar{z}_1$ also vanish
so that the solution reduces to identity. On the other hand,
if $z_1$ is zero, so is $\bar{z}_1$. Then this becomes diagonal situation.

Case(II): In this case, it is quite useful to notice that
the Eqs.(1221-2112, 2332-3223, 2231-3122, 1322-2213,1331-3113) which are
relevant for the diagonal solutions remain intact.
So the diagonal part for this case is the same as the diagonal
solutions. Now it is sufficient to find out the $X(u), Y(u)$ and $Z(u)$.
We use Eqs.(1112, 1121) for 6 unknowns
to determine $X(u)$. Eq.(1121) leads to
\begin{equation}
X(u)=\frac{x_1}{2 ((b,c))} (be+B(u) \bar{e} b).
\end{equation}
Eq.(1112) leads to
\begin{equation}
X(u)=\frac{x_1}{b_1 e c+2 ((c,e))} (B(u) (cc-bb)- e e).
\end{equation}
To determine $Y(u)$, Eqs.(2333 or 3233) can be used. Eq.(2333) leads to
\begin{equation}
Y(u)=\frac{y_1}{2 ((b,c))} (b \bar{e}+G(u) e b).
\end{equation}
Eq.(3233) leads to
\begin{equation}
Y(u)=\frac{y_1}{g_1 \bar{e} c+2 ((c,\bar{e}))} (G(u) (cc-bb)- \bar{e} \bar{e}).
\end{equation}
Above two different expressions for $X(u),Y(u)$ give the same result.

For $b_1=g_1=0$, $X(u), Y(u)$ reduce to the following.
\begin{equation}
X(u)=  {{\left( -1 + {e^u} \right) \,\left( 1 + {e^u} \right) \,{x_1}}\over
    {2\,{e^u}}},
\label{XII1}
\end{equation}
\begin{equation}
Y(u)=   {{\left( -1 + {e^u} \right) \,\left( 1 + {e^u} \right) \,{y_1}}\over
    {2\,{e^u}}}.
\label{YII1}
\end{equation}
For non-trivial diagonal solutions, $X(u), Y(u)$ become
\begin{equation}
X(u)= { {\left( \pm i + e^{3 \eta} \right)
   \left( -1 + {e^u} \right)  \left( 1 + {e^u} \right) x_1 }
 \over {2 {e^u} \left( \pm i + {e^{3 \eta + u}} \right) } },
\label{XII2}
\end{equation}
where $\pm$ corresponds to two non-trivial diagonal solutions in
Eq.(\ref{diagonal}).
\begin{equation}
Y(u)= { {\left( \pm i + e^{3 \eta} \right)
  \left( -1 + {e^u} \right)  \left( 1 + {e^u} \right)  { y_1}}
  \over {2 \left( \pm i + {e^{3 \eta + u}} \right) }}.
\label{YII2}
\end{equation}

To determine the possible conditions on $x_1$ and $y_1$,
we use Eq.(1232) for the 2 unknowns. It gives the following constraints.
\begin{equation}
y_1=e^{-\eta} x_1,
\label{condII1}
\end{equation}
for Eqs.(\ref{XII1},\ref{YII1}) and
\begin{equation}
y_1=\pm i e^{-2 \eta} x_1,
\label{condII2}
\end{equation}
for Eqs.(\ref{XII2},\ref{YII2}) irrespective of the chosen $b_1,g_1$
in Eq.(\ref{diagsol}).
So there are four combinations to consider, which we denote $(++), (+-), (-+),
(--)$, where the first sign refers to $b_1,g_1$ and the second sign
refers to $y_1$.

Let us turn to $Z(u)$.
Eqs.(1123, 1132, 1233, 2133) give the same result for
each different choices of $y_1$ and $b_1, g_1$.
For the trivial diagonal solutions, $Z(u)$ becomes
\begin{equation}
Z(u)= {{\left( -1 + {e^u} \right) \,\left( 1 + {e^u} \right) \,
      \left( {e^{2\,\eta}} + {e^{2\,u}} \right) \,{ z_1}}\over
    {2\,{e^{2\,u}}\,\left( 1 + {e^{2\,\eta}} \right) }},
\label{ZII1}
\end{equation}
for the $y_1$ in Eq.(\ref{condII1}).
For the non-trivial diagonal solutions, $Z(u)$ becomes
\begin{equation}
Z(u)=   {{\left( -1 \pm i {e^\eta} + {e^{2 \eta}} \right)
      \left( {e^\eta} \mp i {e^u} \right)  \left( -1 + {e^u} \right)
      \left( 1 + {e^u} \right)  { z_1}}\over
    {2 {e^u} \left( \pm i + {e^{3 \eta + u}} \right) }},
\label{ZII21}
\end{equation}
for the $(++), (--)$ choices and
\begin{equation}
Z(u)= {{\left( \mp i + e^\eta \right) \left( -1 \pm i e^\eta + e^{2 \eta}
\right)
     \left( {e^\eta} \pm i\,{e^u} \right) \left( -1 + {e^u} \right)
     \left( 1 + {e^u} \right) { z_1}}\over
 {2 e^u \left( \pm i + {e^\eta} \right) \,\left( \pm i + e^{3 \eta + u} \right)
}}
\label{ZII22}
\end{equation}
for the $(+-), (-+)$ choices.

We need to determine the possible conditions on $z_1$.
We use Eqs.(1223 or 2132). For $Z(u)$ in Eq.(\ref{ZII1}), there does not
exist consistent solution for $z_1$, {\it ie.} $z_1,x_1$ must be zero.
For the $(++), (--)$ choices,
\begin{equation}
z_1={ {\left( \mp i + e^{\eta} \right) \left(1 \mp i e^{\eta} - e^{2 \eta}
\right) {x_1}^2}
 \over {2 e^{3 \eta} \left( \pm i + e^\eta \right) } }.
\end{equation}
For the $(+-), (-+)$ choices, it turns out there is no non-trivial
solution for $z_1$, {\it ie.} $z_1, x_1$ must vanish.
This is the end of the Case(II). The result is the following for
the sign choices $(++), (--)$.
\begin{eqnarray}
X(u) &=& {{ (\mp i + e^\eta ) ( -1 \pm i e^\eta + e^{2 \eta} )
 (-1 + e^u) (1 + e^u ) x_1}\over {2 e^u ( \pm i + e^{3 \eta + u} ) }}, \non  \\
Y(u) &=& {{ (1 \pm i e^\eta ) ( -1 \pm i e^\eta + e^{2 \eta} )
 (-1 + e^u) (1 + e^u ) x_1}\over {2 e^{2 \eta} (\pm i +e^{3 \eta + u})}}, \non
\\
Z(u) &=& {{ e^{-3 \eta - u} \left( \mp i + e^\eta \right)
 {{ ( -1 \pm i e^\eta + e^{2 \eta} ) }^2}
    (1 - e^u) (e^\eta \mp i e^u) (1 + e^u) x_1^2 }\over
    {4 ( \pm i + e^\eta ) ( \pm i + e^{3 \eta + u} ) }},  \\
B(u) &=& {{ (e^{3 \eta} \pm i e^u) }\over {e^u ( \pm i + e^{3 \eta + u} ) }},
 \non \\
G(u) &=& {{ e^u ( e^{3 \eta} \pm i e^u ) }\over {(\pm i + e^{3 \eta + u}) }}.
 \non
\end{eqnarray}
The solutions which we found satisfies all $9 \times 9$ equations.

Case(III): This case can be solved in the same way as in the case (II) since
the equations for this case is exactly the same.
We have only to replace every unbarred variables with barred ones.
\begin{eqnarray}
\bar{X}(u) &=& {{ (\mp i + e^\eta ) ( -1 \pm i e^\eta + e^{2 \eta} )
 (-1 + e^u) (1 + e^u ) \bar{x_1} }\over {2 e^u ( \pm i + e^{3 \eta + u} ) }},
 \non   \\
\bar{Y}(u) &=& {{ (1 \pm i e^\eta ) ( -1 \pm i e^\eta + e^{2 \eta} )
 (-1 + e^u) (1 + e^u) \bar{x_1}}\over {2 e^{2 \eta} (\pm i +e^{3 \eta + u})}},
 \non  \\
\bar{Z}(u) &=& {{ e^{-3 \eta - u} \left( \mp i + e^\eta \right)
 {{ ( -1 \pm i e^\eta + e^{2 \eta} ) }^2}
    (1 - e^u) (e^\eta \mp i e^u) (1 + e^u) {\bar{x_1}}^2 }\over
    {4 ( \pm i + e^\eta ) ( \pm i + e^{3 \eta + u} ) }},  \\
B(u) &=& {{ (e^{3 \eta} \pm i e^u) }\over {e^u ( \pm i + e^{3 \eta + u} ) }},
 \non  \\
G(u) &=& {{ e^u ( e^{3 \eta} \pm i e^u ) }\over {(\pm i + e^{3 \eta + u}) }}.
 \non
\end{eqnarray}

\newpage
\section*{IV. Discussions}
The most general solutions of $K$-matrix to be determined
from reflection equation for known $R$-matrix are highly desired.
Obviously it would be best to obtain them in algebraic way.
But until now, only small number of solutions are known for small
size $R$-matrices in a direct way as in the present paper.

We solved the reflection equation for the quantum Mikhailov-Shabat
model. This model is known to have two non-trivial diagonal solutions
as well as trivial diagonal solution.
The most general solutions which we found consist of three families
corresponding to each diagonal solutions.
One family of solutions depends on two arbitrary parameters.
If one of the parameters vanishes, the other must also vanish
so that the solutions reduces to trivial identity solution.
The other two families for each non-trivial diagonal solutions
have only one arbitrary parameter.

This solution can be used to compute the Hamiltonian including
the boundary terms for the corresponding open spin chains
along the lines in Ref.\cite{Skl}. It can also be used to calculate
the boundary $S$-matrix of the model along the lines in Ref.\cite{GZ}
using the $K$-matrix and the bulk $S$-matrix.
The bulk $S$-matrix for this model was obtained in Ref.\cite{Smi} from
the $R$-matrix.

On the other hand, the quantum Mikhailov-Shabat model has
Lagrangian description in terms of single scalar field.
The classical boundary
$S$-matrix can also be computed in the way pursued in Ref.\cite{CDRS}.
It would be nice if we can see whether they give the consistent result.

\section*{Acknowledgement}
I would like to thank Ed Corrigan.
I am also grateful to Roberto Tateo and Alistair Macintyre for discussions.
This work was supported by Korea Science and Engineering Foundation through
foreign post-doctoral program and in part by University of Durham.

\end{document}